\documentclass[aps,prb,twocolumn,showpacs,floatfix,superscriptaddress,amssymb,amsmath]{revtex4}
\usepackage{graphicx}
\usepackage{dcolumn}
\usepackage{bm}
\usepackage{color}
\newcommand{\be}{\begin{equation}}
\newcommand{\ee}{\end{equation}}
\newcommand{\bea}{\begin{eqnarray}}
\newcommand{\eea}{\end{eqnarray}}

\newcommand{\s}{\sigma}
\newcommand{\vare}{\varepsilon}

\newcommand{\re}{\mbox{e}}
\newcommand{\ba}{\begin{array}}
\newcommand{\ea}{\end{array}}
\def\nn{\nonumber\\}
\def\vare{{\varepsilon}}
\def\up{\uparrow}
\def\down{\downarrow}

\begin{document}

 \title{Rashba spin-orbit interaction in graphene and zigzag nano-ribbons }

 \author{ Mahdi Zarea and Nancy Sandler} 
 \author{} 
 \author{}
 \affiliation{Department of Physics and Astronomy, Nanoscale
   and Quantum Phenomena Institute, and Condensed Matter and Surface Science Program,\\Ohio University, Athens, Ohio
   45701-2979}

 \date{\today}

 \begin{abstract}

We investigate the effects of Rashba spin-orbit (RSO) interactions on the electronic band-structure and corresponding wavefunctions of graphene. By exactly solving a tight-binding model Hamiltonian we obtain the expected splitting of the bands -due to the SU(2) spin symmetry breaking- that is accompanied by the appearance of additional Dirac points. These points are originated by  valence-conduction band crossings. By introducing a convenient gauge transformation we study a model for  zigzag nanoribbons with RSO interactions. We show that RSO interactions lift the quasi-degeneracy of the edge band while introducing a state-dependent spin separation in real space. Calculation of the average magnetization perpendicular to the ribbon plane suggest that RSO could be used to produce spin-polarized currents. Comparisons with  the intrinsic spin-orbit (I-SO) interaction proposed to exist in graphene are also presented.
 \end{abstract}

\pacs{ 73.20.At,85.75.-d, 73.63.Bd, 81.05.Uw, 73.43.f} 
\maketitle

\section{Introduction}

Understanding the mechanism for the generation and manipulation of spin polarized currents is one of the 
greatest challenges for the development of spin-based devices. Much of the advance in the field in latest 
years \cite{Murakami-adv} has been achieved by studying semiconductor materials which make up the bulk of 
current electronic circuitry. Among the mechanisms proposed to induce spin-polarized currents, the spin Hall effect (SHE) appears as the most efficient one. The SHE refers to the phenomenon in which a spin-polarized current is created when an external bias voltage is applied to the system. The effect is based on a coupling between spin and momentum degrees of freedom, and usually the existence of some kind of spin-orbit (SO) interaction in the particular system under study is invoked. 
For bulk semiconductor materials for example, the SO interaction, has been proposed to lead to two different manifestations of SHE:
a) the {\it intrinsic}  SHE \cite{Murakami,Sinova}, in which the material inherits a strong SO interaction from its atomic constituents or due to its crystalline symmetries (lack of inversion symmetry) and b) the {\it extrinsic} SHE \cite{Dyakonov,Hirsch}, in which spin-polarized currents appear as a consequence of electron scattering by a SO dependent scattering potential. In this last situation the scattering potential may be caused for instance, by magnetic impurities that couple via a spin-orbit term to the conducting electrons or by defects that produce spin-dependent scattering. 
Among  these  scenarios one possibility is when interfaces or surfaces are considered. In this case,  the existence of the interface/surface introduces inversion symmetry breaking and thus, materials that do not fall into the categories cited above can also exhibit SHE. The effective SO interaction generated in this situation is known as the Rashba spin-orbit (RSO) interaction  responsible for the Rashba effect\cite{Rashba}.
Among the many materials in which RSO interactions could be exploited to obtain spin-polarized currents, graphene presents a unique and intriguing case. The material, first isolated as a single layer of graphite in 2004 \cite{Gaim}, gives access to a crystalline surface with linear dispersion around two independent points in its Brillouin zone, the Dirac points. The special dispersion plus its crystal structure (two-atom base triangular lattice) makes possible to calculate low-energy properties using Dirac-type models as the vast literature in recent years shows\cite{Antonio}. Furthermore, the two sublattices of the honeycomb structure makes appropriate the use of two-valued wave-functions, or spinors, for calculations of various properties. The relativistic description for low energy properties has also been used to argue for the relevance of additional terms in the standard Dirac Hamiltonian, including an intrinsic spin-orbit (I-SO) interaction  represented by a second-neighbor spin-dependent hopping term that respects all the symmetries of the graphene plane\cite{Kanemele}. One of the consequences of the I-SO interaction is to make possible the existence of spin-polarized edge states in a new phase of matter, the quantum Spin Hall (QSH) phase\cite{Kanemele2}. In previous works we have studied the physics introduced by this interaction in narrow graphene ribbons with armchair and zigzag edge terminations and in the presence of electron-electron interactions\cite{Zarea1, Zarea2}. We have shown that the I-SO interaction does not change the metallic behavior of armchair nanoribbons in contrast with the predicted result for graphene sheets. Moreover, the interaction produces spin-filtered states localized along the edges of the ribbon, (independent of edge termination) and, as a consequence, the current induced by an applied low external voltage is spin polarized. These results are in good agreement with several numerical and analytic studies\cite{Son1,Son2,Wimmer,Abanin,Kim,Wonsch,Huang,Kuma,Jiang,Rado} that point to various magnetic instabilities  that graphene ribbons may sustain, leading to some kind of magnetic order along the edges. Unfortunately, all numerical estimates for the strength of the I-SO interaction remain, although still controversial, rather small, in the range of $0.05-0.0011meV (600-13mK)$ \cite{Trickey}. 

However, as a pure two-dimensional material, a graphene flake on a substrate lacks inversion symmetry and it is natural to expect that a RSO interaction may introduce important changes to the material properties. The RSO coupling $\lambda_R$ is controlled by the applied bias and although predicted to be in the range of $\lambda_R \sim 1 meV$ (Refs. \cite{Boet,Min,Guinea}) recent experiments have shown that it can reach values up to $\lambda_R \sim 200 meV$ for graphene deposited on a $Ni$ substrate \cite{Dedkov}. Furthermore, experiments also have shown short spin relaxation times that suggest an important effect of spin-orbit interactions in graphene
\cite{Tombros,vanWees,Hernando,Vary}.

These new developments highlight the need for a better understanding of the role played by the RSO interaction on various properties of graphene and graphene ribbons. 
In this work we address the questions raised by the presence of RSO in graphene sheets and zigzag ribbons. The RSO interaction strongly affects  the dispersion relation near the two-independent Dirac points as well as the nature of the corresponding wave-functions as we will show below. As a SU(2) breaking symmetry interaction, it favors a spatial spin ordering but in contrast to the I-SO interaction introduced above, the spin order is not originated on each individual state having the same spatial spin distribution but it emerges from averaging over several states.

\section{Graphene sheet: model}

\begin{figure}[!]
 \includegraphics[width=.4\textwidth]{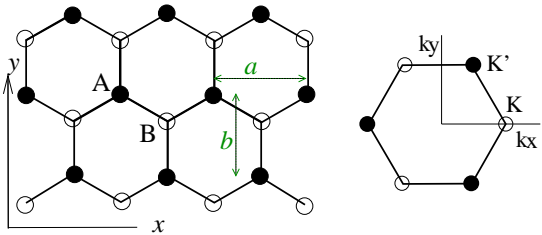}
 \caption{Left panel: graphene lattice. Lengths along the $x$ and $y$ directions 
 are measured in units of $a$ and $b=a\sqrt{3}/2$ respectively. 
Right panel: First Brillouin
zone. $K=(4\pi/3a,0)$ and $K'=(2\pi/3a,\pi/b)$ are Dirac points.
}
 \label{gra}
\end{figure}

To describe a graphene sheet (an infinite mono-layer of carbon atoms arranged in a honeycomb structure), we introduce, as usual, two sublattices $A$ and $B$ with their respective atoms connected by vectors:
\bea
&&{\bm \delta}_1 = a(0,1/\sqrt{3})
\nn &&{\bm \delta}_2 = a(1/2,-1/2\sqrt{3})
\nn &&{\bm \delta}_3 = a(-1/2,-1/2\sqrt{3})
\eea
where $a=2.4~\AA$ is the lattice constant [see Fig.~(\ref{gra})]. 

In the absence of spin-orbit  interactions spin-up and spin-down electrons
are degenerate. The SU(2) symmetric Hamiltonian and four-component spinor wave-function in momentum space are given by:
\be
H  =  \left( \begin{array}{cccc}
 0 & \varphi & 0 & 0 \\
 \bar\varphi & 0 & 0 & 0\\
0 & 0 & 0 & \varphi\\
0 & 0 &\bar\varphi & 0
\end{array}\right)
~~~~%
\Psi = \left(\begin{array}{c} 
    u_{A\up}         \\
    u_{B\up}        \\
    u_{A\down}\\
    u_{B\down}   
\end{array}\right)
\label{eq:hamiltonian}
\ee
with $\varphi(k_x,k_y) =  t( \re^{ik_y2b/3}+2\cos{\frac{k_xa}{2}}\re^{-ik_yb/3})$. In these expressions $\bar\varphi$ is defined as 
$\bar\varphi(k_x,k_y)=\varphi(k_x,-k_y)$. Notice that for real values of $k_y$, $\bar\varphi = \varphi^*$. The eigenvalues of Eq.(\ref{eq:hamiltonian}) are 
$E=\pm\vare=\pm\sqrt{\varphi\bar\varphi}$ 
and the corresponding eigenvectors are defined in terms of the angle $\alpha_0$ as:
\be
\Psi_{\pm\up}=N\left( \begin{array}{c}
 \re^{i\alpha_0/2}  \\
 \pm \re^{-i\alpha_0/2}\\
 0  \\
 0\\
\end{array}\right)\re^{ik_xx}\re^{ik_yy}
\ee
\be
\Psi_{\pm\down}=N\left( \begin{array}{c}
 0  \\
 0\\
 \re^{i\alpha_0/2}  \\
 \pm \re^{-i\alpha_0/2}\\
\end{array}\right)\re^{ik_xx}\re^{ik_yy}
\label{spinor}
\ee
with $\varphi=|\varphi|\re^{i\alpha_0}$ and $N$ the normalization factor. For neutral graphene,  $\Psi_+$ ($\Psi_-$) represents solutions with $E > 0$  $(E < 0)$ and refers to electron (hole) conduction (valence) bands. In this language, the particle-hole symmetry implies that for each electron state with energy $E=\vare$ and eigenstate characterized by  $\alpha_0$, there is a 
hole state with $E=-\vare$ and eigenstate given by $\alpha_0+\pi$. 

\subsection{ Rashba spin-orbit interaction}

Depositing graphene on substrates and/or applying external fields makes possible to 
introduce a controllable RSO interaction. In the following we take the effective electric field $\mathcal{E}$ perpendicular to the graphene plane \cite{Martino-prl,Kanemele2,Yao}. The Rashba Hamiltonian is then given by:
\be
H_R=\sum_{<ij>}ic^{\dag}_i(\vec u_{ij}\cdot\s)c_j+h.c
\ee
where $\s$ represents the Pauli spin operator for the spin degree of freedom.
Here $\vec u_{ij}$  is given by:  
\be
\vec u_{ij}={e\over2m^2dv_F}\vec{\mathcal{E}}\times\vec{\delta}_{ij}=-{\lambda_R\over d}
\hat z\times\vec{\delta}_{ij}.
\ee
where $\vec{\mathcal{E}}$ is the applied electric field in the direction perpendicular to the graphene sheet, $d = a/\sqrt{3}$ is the distance between the two adjacent sites $(i, j)$ and $\vec{\delta}_{ij} = {\bm \delta_j} - {\bm \delta_i}$ is a vector on the graphene plane. 

With these definitions the  RSO interaction takes the form:
\be
H_R=i c^{\dag}_{A\alpha}R^{\alpha\beta}c_{B\beta}+h.c
\ee where
\bea
&&R^{\xi\eta}=(\re^{ik_yd}\s^{\xi\eta}_x
+\re^{ik_xd\sqrt{3}/2-ik_yd/2}(-\s^{\xi\eta}_x/2-\s^{\xi\eta}_y\sqrt{3}/2)\nn
&&+\re^{-ik_xd\sqrt{3}/2-ik_yd/2}(-\s^{\xi\eta}_x/2+\s^{\xi\eta}_y\sqrt{3}/2)
\eea
where $(\xi, \eta)$ stand for spin up and down.

The RSO interaction couples spin-up and spin-down states, breaking the corresponding SU(2) symmetry, leading to the Hamiltonian (written in the
four-component spinor $\Psi$ basis):
\be
H = \left(\begin{array}{cccc} 
    0   & \varphi_0   &  0        & i\varphi_+         \\
    \bar\varphi_0  & 0    & -i\bar\varphi_- &  0        \\
   0        &   i\varphi_- & 0   & \varphi_0  \\
   -i\bar\varphi_+   &   0        & \bar\varphi_0    & 0   
\end{array}\right)~~~
\Psi = \left(\begin{array}{c} 
    u_{A\up}         \\
    u_{B\up}        \\
    u_{A\down}\\
    u_{B\down}   
\end{array}\right).\label{Hamiltonian-wave}
\ee
where
\bea
&&\varphi_0=t\re^{i2k_yb/3}(1+2\re^{-ik_yb}\cos(k_xa/2))\nn
&&\varphi_+=\lambda_R\re^{i2k_yb/3}(1+2\cos(k_xa/2+2\pi/3)\re^{-ik_yb} )\nn
&&\varphi_-=\lambda_R\re^{i2k_yb/3}(1+2\cos(k_xa/2-2\pi/3)\re^{-ik_yb} ).
\eea

The eigenvalue equation is given by:
\be
E_+^2E_-^2=\Phi\bar\Phi
\ee
in which we have defined
\bea
&&E^2_{\pm}=E^2-\vare_0^2-\vare_{\pm}^2\nn
&&\Phi=i\varphi_+\bar\varphi -i\varphi\bar\varphi_-=E^+E^- \re^{i\nu}\label{eigenvalues}
\eea
and $\vare_{0; \pm}=|\varphi_{0; \pm}|$.  The angle $\nu$ is defined by
\be
\tan \nu={\Im\Phi}/{\Re\Phi}.
\ee
Note that $\nu$ is {\it independent} of the RSO coupling.

Equation (\ref{eigenvalues}) shows explicitly that $E\to-E$, representing the particle-hole symmetry, is preserved by the RSO interaction. Using this property, in the rest of the paper we will focus on conduction bands only.

\begin{figure}[!]
 \includegraphics[width=.5\textwidth]{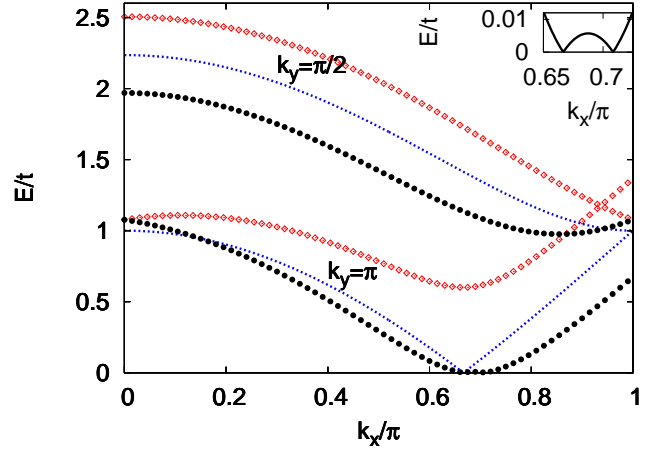}
 \caption{
Energy bands as function of $k_x$ of an infinite graphene plane without
(small dotted lines) and with (filled circles and empty diamonds) RSO. The strength of the RSO interaction is $\lambda_R=0.2t$. The plotted bands correspond to $k_y=\pi$ (lower set) and $k_y=\pi/2$ (upper set). The inset shows the new lower energy band with $k_y=\pi$ and the splited 
Dirac point. A finite gap separates all other conduction and valence bands. }
 \label{ra-bulk}
\end{figure}

\begin{figure}[!]
 \includegraphics[width=.5\textwidth]{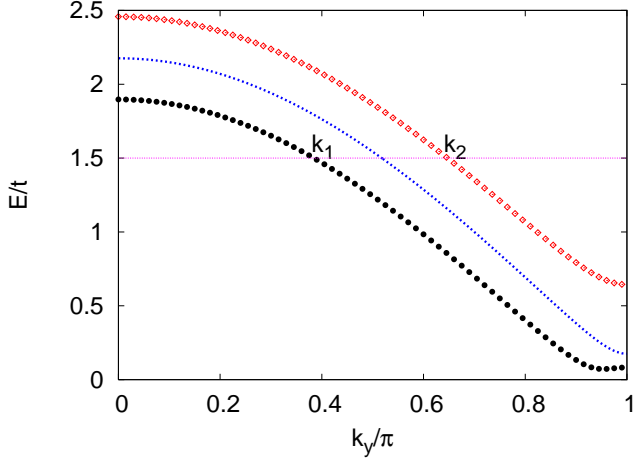}
 \caption{
Energy bands as function of $k_y$ of an infinite graphene plane without (dotted lines) and with (full lines). The strength of the RSO interaction is $\lambda_R=0.2t$. The plotted bands correspond to $k_x=0.6\pi$. 
As described in the text there are 
four degenerate states at $k_y=\pm k_1$ and $k_y=\pm k_2$ for each $k_x$ value.   }
 \label{ra-bulk-y}
\end{figure}

In Fig.~(\ref{ra-bulk})  and Fig.~(\ref{ra-bulk-y}) we plot few bands for the infinite graphene plane in
the presence of the RSO interaction. Two new features appear: 
a) due to the breaking of the SU(2) symmetry, each of the degenerate bands (in the absence of spin orbit interaction) splits into two. For a given
$(k_x, k_y)$ the energies of the newly separated upper ($E_u$) and lower bands ($E_l$) are related by $E_u^2-\vare_0^2-\vare_{\pm}^2 =-(E_l^2-\vare_0^2-\vare_{\mp}^2)$. 
b) The inset of Fig.~(\ref{ra-bulk}) shows the change in the band with $k_y=\pi$ which originally 
touches the corresponding valence band at a Dirac point located at $K=(2\pi/3a, \pi/b)$. 

\begin{figure}[!]
 \includegraphics[width=.5\textwidth]{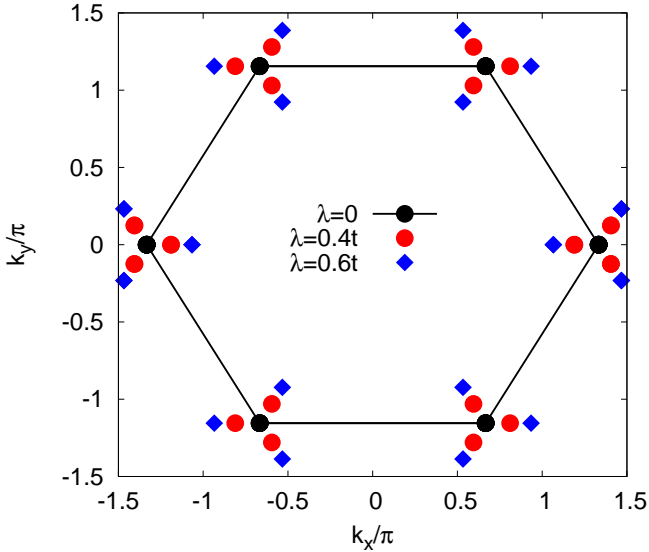}
 \caption{ The positions of new zero-energy (Dirac) points in momentum space for different values of the RSO interaction strength $\lambda_R$. The distribution of the new Dirac points around the original ones has the $2\pi/3$ rotational symmetry of the graphene lattice.
 }\label{dirac}
\end{figure}

A remarkable consequence of the RSO interactions is the splitting of the original Dirac point caused by crossings of conduction and valence bands. The location of the new points in reciprocal space respects the underlying honeycomb symmetry and depends on the strength of the interaction $\lambda_R$ as shown in Fig.~(\ref{dirac}). For the Dirac point at $K = (2\pi/3; \pi/b)$ shown in Fig.~(\ref{ra-bulk}), the position of one new point is at $(k'_x, \pi/b)$ with $k'_x$ given by:
\be
\cos(k'_x/2)=\frac{1}{2}\frac{t^2-2\lambda_R^2}{t^2+\lambda_R^2}.
\ee
In the linear (Dirac) approximation of the Hamiltonian and for small values of the interaction strength $\lambda_R$, the splitting of the Dirac points is missed, and the low-energy effective Hamiltonian describes
graphene with RSO interactions as a zero-gap semiconductor as reported in previous works \cite{Kanemele}.

It is important to remark that the RSO interaction does not open a gap in the spectrum at the Dirac point, in contrast to the I-SO interaction mentioned above. 

To solve for the eigenstates of the Hamiltonian (\ref{Hamiltonian-wave}) we notice
first that in the limit $\lambda_R\to 0$ the  spinor introduced in Eq.(\ref{Hamiltonian-wave})
takes the form $\Psi=\re^{i\nu/2}\psi_\up +\re^{-i\nu/2}\psi_{\down}$
where $\psi_{\up\down}$ are the two degenerate spinors defined in the absence of the RSO 
interaction. The four components of $\Psi$ satisfy the following relations:
\bea
&&\frac{u_{A\up}}{u_{A\down}}=\frac{E_-}{E_+}\re^{i\nu}
\nn&&\frac{u_{B\up}}{u_{B\down}}=\frac{E_+}{E_-}\re^{i\nu}
\nn&&u_{B\down}=\frac{\bar\varphi}{E}u_{A\up}-i\frac{\bar\varphi_-}{E}u_{A\up}
\eea  
which lead to the eigenstates:
\be
\Psi=\left(\begin{array}{c}
u_{A\up}\\
u_{B\up}\\
u_{A\down}\\
u_{B\down}\end{array}\right)=
\left(\begin{array}{c}
\re^{i\nu/2}\left(\begin{array}{c}
                                   \sqrt{E^-\over E^+}\re^{i\alpha/2}\\
                                   \sqrt{E^+\over E^-}\re^{-i\alpha/2}\end{array}\right)\\
\re^{-i\nu/2}\left(\begin{array}{c}
                                   \sqrt{E^+\over E^-}\re^{i\alpha/2}\\
                                   \sqrt{E^-\over E^+}\re^{-i\alpha/2} \end{array}\right)
\end{array}\right).\label{wave-bulk}
\ee
Here $\alpha$ is defined by:
\bea
\re^{i\alpha}&=&\frac{\varphi}{E}\frac{E^+}{E^-}+i\frac{\varphi_+}{E}\re^{-i\nu}
\nn &=& \frac{\varphi}{E}\frac{E^-}{E^+} +i\frac{\varphi_-}{E}\re^{i\nu}.
\eea
In the limit $\lambda_R\to0$, $\sqrt{E^+\over E^-}\to 1$
and $\alpha\to\alpha_0$. The expressions above (Eq.(\ref{wave-bulk})) correspond to a state with energy $E$ in the split upper band and momentum $(k_x,k_y)$. The sate with the same momentum components in the split lower band is obtained by the replacement $\nu\to \nu+\pi$. The remaining particle-hole symmetric states (in the split valence bands) are obtained by taking $\alpha\to \alpha+\pi$.

Wave function (\ref{wave-bulk}) has the property of $u_{A\up}={\bar u_{B\down}}$
and $u_{B\up}={\bar u_{A\down}}$. This reflects that, in the presence of the SU(2) symmetry breaking RSO interaction the probability of finding an electron in the spin up state i.e $|u_{A\up}|^2+|u_{B\up}|^2$ is equal to the probability of finding it in the spin down state i.e  $|u_{A\down}|^2+|u_{B\down}|^2$. This is a direct consequence of the fact that the  RSO interaction does not break time reversal symmetry.
However, as we will see below, this dos not exclude the possibility of
separating the spin-up and spin-down  electron states and localizing them at different positions  in the sample while preserving a zero net magnetization.

\section{Zigzag graphene nanoribbons}

\subsection{Zigzag nanoribbons without Rashba spin-orbit interaction}

\begin{figure}[!]
 \includegraphics[width=.5\textwidth]{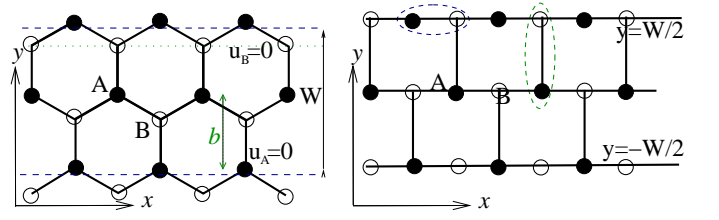}
 \caption{Hard-wall boundary conditions for ZGR imposes $u_A=0$ on the lower edge and $u_B=0$ just before the upper edge (green dotted line).
The deformed lattice shown on the right side, corresponds to a gauge transformation (see text) and it is equivalent of labeling both $A$ and $B$ sites in each zigzag line with the same $y$-coordinate. 
The horizontal oval shows our choice of the unite cell and the vertical oval shows the choice of the unit cell which corresponds to Eq.(\ref{cphi}). 
}
 \label{ZGR}
\end{figure}

To study the interplay between confinement and the RSO interaction we analyze the case of zigzag graphene (ZGR) nanoribbons, defined according to Fig.~(\ref{ZGR}).

As standard practice the hard-wall boundary conditions are imposed by setting $u_A(y=0)=0$  on the lower border and $u_B(W-b/3)=0$ on the line located at a distance $b/3$ below the upper border \cite{Fujita,Nakada,Brey,Hikihara,Beenakker3}.
After close inspection, however, one realizes that it is more convenient to label all the atoms alone each zigzag line with the same $y$-coordinate. This is equivalent to work with the deformed lattice shown in the right panel of Fig.~(\ref{ZGR}) and it amounts to perform a global gauge transformation $c_B(k_y)\to c_{B}(k_y)\re^{2ik_yb/3}$ on the original Hamiltonian. It also represents a different choice of unit cell as described in Fig.~(\ref{ZGR}).
For symmetry reasons we also set the $y$-axis to be in the center of the ribbon. With this choice, the  hopping term $\varphi$ reduces to:
\be
\varphi(k_x,k_y) =  t( \re^{ik_yb}+2\cos{\frac{k_xa}{2}})\label{dphi}
\ee
and the boundary conditions are:
\be
u_A(y=-W/2)=0, ~~u_B(y=W/2)=0. \label{dbc}
\ee

Notice that in most of the literature on graphene ribbons the usual convention for
$\varphi$ and the boundary conditions are:
\bea
&&u_A(y=-W/2)=0, ~~u_B(y=W/2-1)=0
\nn&&  \varphi(k_x,k_y) =  t( 1+2\cos{\frac{k_xa}{2}}\re^{-ik_yb})\label{cphi}
\eea
which correspond to choosing a unit cell along the vertical link in the right-side panel of Fig.~(\ref{ZGR}).

The wave-function of the ZGR can be found in a straightforward manner as follows.
Since  $k_x$ is a good quantum number, the wave-function for a given $k_x$ must be a superposition of degenerate states with different $k_y$ values. 
In the absence of SO there are only two degenerate spinors for  each $k_x$ namely
at $k_y = k$ and $k_y = -k$. Therefore the wave-function is the superposition of these
two spinors: $\Psi=a\Psi(k_x, k)+b\Psi(k_x, -k)$. After applying
the boundary conditions as given in Eq.(\ref{dbc}), we find  $b=-a$ such that
\bea
\Psi&=&N\re^{ik_xx}\left( \begin{array}{c}
 \sin(\alpha_0/2+ky-n\pi/2)  \\
 \sin(-\alpha_0/2+ky-n\pi/2)\end{array}\right)
\eea
where $k$  satisfies
\be
\alpha_0-kW=n\pi.\label{zgr-cnd}
\ee

\begin{figure}[!]
 \includegraphics[width=.5\textwidth]{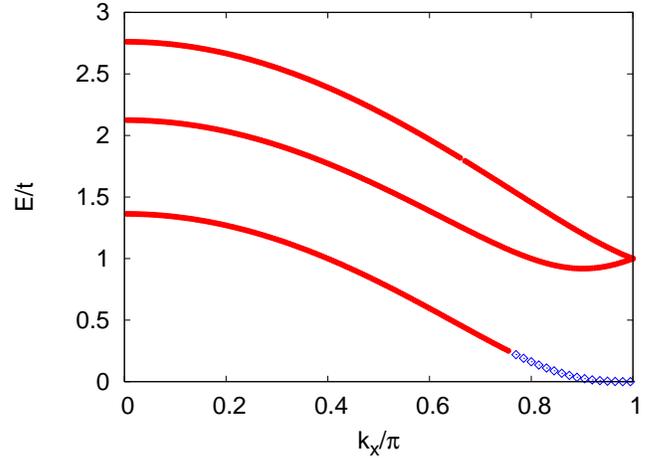}
 \caption{Energy bands of a zigzag ribbon with $W=4b$ 
in the absence of the RSO interaction. Each band is doubly 
degenerate due to the SU(2) spin symmetry. The edge band (diamonds)
corresponds to an imaginary value for the label $k$ in Eq.(\ref{zgr-cnd}). }
 \label{zgr-band}
\end{figure}

Figure (\ref{zgr-band}) shows the conduction bands of a ribbon with $W=4b$. Zigzag ribbons present two remarkable features as compared to graphene sheets: the momentum across the ribbon $k_y$ can take complex values between two Dirac points \cite{Fujita,Nakada,Brey,Beenakker3,Hikihara}, producing edge states and their band-structure depends on the width $W$ or the number of zigzags chains $N = W/b -1$. It can be shown \cite{Beenakker,Beenakker2,Li,ZS3} that in zigzag ribbons with odd number of chains $N$, the so-called 'zigzag/zigzag' configuration, conduction and valence edge bands cross at $k_x a= \pi$. In contrast, ribbons with even number of chains $N$, in the 'zigzag/anti-zigzag' configuration, edge bands do not cross albeit remain degenerate at $k_x a= \pi$.

\subsection{Zigzag nanoribbons with Rashba spin-orbit interaction}

As seen in the previous section, ZGRs present the peculiar feature of edge states which
remain highly quasi-degenerate at low-energies for wide ribbons. 
These states are expected to be strongly affected by the presence of a RSO interaction. Below we proceed to obtain the exact expressions for the band-structure and corresponding eigenstates for the ZGR with RSO interactions. 

It is necessary to remark first that, since the RSO interaction involves nearest neighbor hopping, the boundary conditions as imposed in Eq.~(\ref{dbc}) remain unchanged. However, for a given value of $k_x$, there are four degenerate states at $k_y=\pm k_1$ and  $k_y=\pm k_2$ in contrast with the previous case (with only two degenerate states at $\pm k$). This is easily seen in Fig.~(\ref{ra-bulk-y}).
Notice that there are certain energies, such that it seems that there are
 only two degenerate states, however the wave-function is really the superposition of four spinors with  $\pm k_1$ and $\pm k_2$ taking imaginary or complex values.
The general wave function is:
\be
\Psi^{ZGR}=a\psi(k_1)+ b\psi(-k_1) +c\psi(k_2)+d\psi(-k_2)\label{gen-wave}
\ee
where $\psi(k_i) = \psi(k_x, k_i)$ and $k_1$ and $k_2$ satisfy the condition given by the degeneracy:
\be
E(\pm k_1)=E(\pm k_2).
\ee
Imposing the boundary conditions given in Eq.~(\ref{dbc}) yields
\bea
\frac{E_1^+\sin(\mu_1^++\delta)}
    {E_1^-\sin(\mu_1^-+\delta)}
=\frac{E_2^+\sin(\mu_2^++\delta)}
    {E_2^-\sin(\mu_2^-+\delta)}
\eea
where $\mu_{i}^{\mp}=(\nu_i\pm\alpha_i\mp k_iW)/2$ and
$\delta=\pm\pi/2$.
These two equations define the band structure and the corresponding
wave-function in terms of the width $W$ and the RSO coupling $\lambda_R$.
The wave function coefficients are given by:
\bea
&&a=-b=N\sqrt{E_2^-\over E_2^+}\sin(\mu_2^++\delta)
\nn&&d=-c=N\sqrt{E_1^-\over E_1^+}\sin(\mu_1^++\delta)
\eea
where $N$ is the normalization factor. From these expressions it can be shown that
$\Psi^{ZGR}_{A\up}(y)=i\Psi^{ZGR}_{B\down}(-y)$ and $\Psi^{ZGR}_{A\down}(y)=i\Psi^{ZGR}_{B\up}(-y)$.

\begin{figure}[!]
 \includegraphics[width=.5\textwidth]{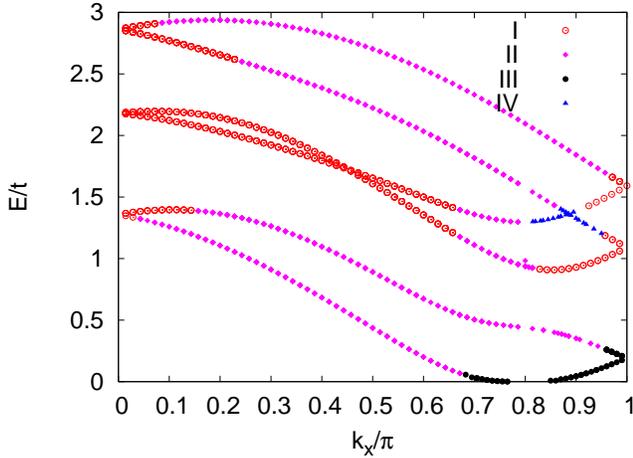}
 \caption{Energy bands of a ZGR with $W=4b$ and $\lambda_R=0.4t$.
Different  regions correspond to values of $k_1$ and $k_2$: I) both real, II) one real and the other one imaginary
III) both imaginary,
IV) complex with $k_1=k_2^*$.}
 \label{band-4-.4}
\end{figure}

Figure (\ref{band-4-.4}) shows the conduction bands for a ribbon with
$W=4b$ and $\lambda_R = 0.4t$. Starting at $k_x=0$ both parameters $k_1,k_2$ are real (region $I$). As $k_x$ is increased,
$k_2$ goes to  $\pi$ or zero and in region $II$ it becomes complex (with
constant real part equal to  $\pi$) or purely imaginary. 
In region $III$ both $k_1$ and $k_2$ take imaginary values with a constant real part  of $\pi$ or zero.
 The energy of the lower band goes to zero $E=0$ at the point $k_x^0$ defined by:
\be
\cos({k_x^0a\over2})=\sqrt{{3\lambda_R^2\over4\lambda_R^2+4t^2}}.\label{kx0}
\ee

Notice finally that there is also a region ($IV$) where $k_1$ and $k_2$ are complex conjugate of each other. As it occurs with the I-SO interaction, the presence of the RSO interaction lifts the apparent quasi-degeneracy of the edge band while preserving the Dirac points\cite{Zarea2}.  The expression for the dispersion of the edge bands of an $N$-wide ribbon is readily obtained and is given by:
\be
E \approx \pm t(k_xa-k_x^0a)^{N}
\label{EvsN}
\ee
where $k_x^0$ is defined in Eq.(\ref{kx0}).

\begin{figure}[]
 \includegraphics[width=.5\textwidth]{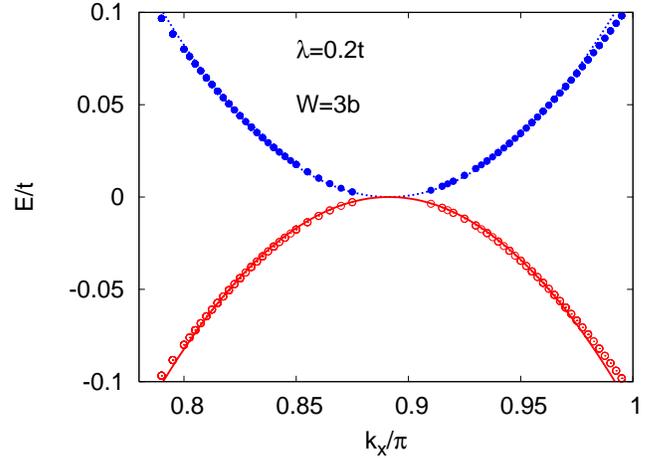}
\includegraphics[width=.5\textwidth]{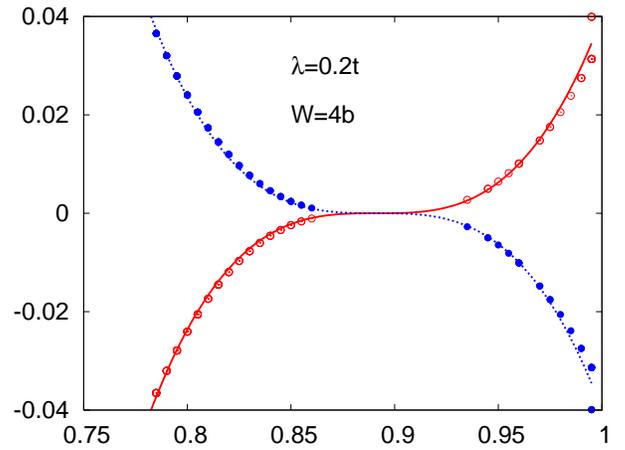}
 \caption{
The upper panel shows the edge bands of a ribbon with $W=3b$ ($N=2)$ and $\lambda_R=0.2t$.
Conduction and valence edge bands do not cross at the band center. In contrast, the lower panel shows a $W=4b$ ($N=3$) ribbon with crossing bands. The full lines are fits using the expression in Eq.(\ref{EvsN}).
}
 \label{ENW3}
\end{figure}

It is interesting to notice that RSO
interactions preserve the power law energy dispersion and edge bands crossing/anticrossing feature as shown in  Figs.~(\ref{ENW3}).

\begin{figure}[!]
 \includegraphics[width=.5\textwidth]{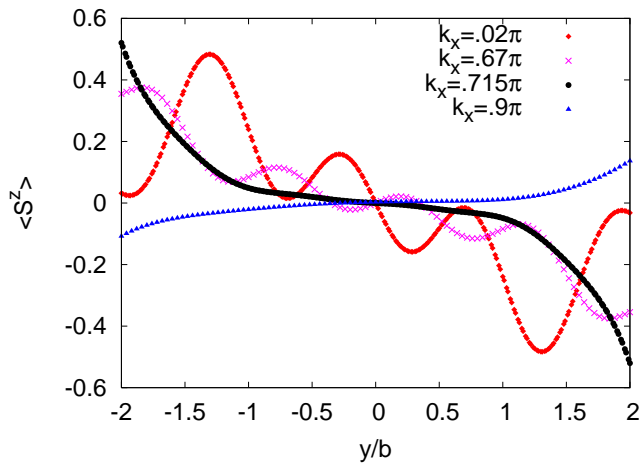}
 \caption{
Expectation value of the $z-$ projection of the spin operator, $<S^z> $ as a function of position across the ribbon. The curves are calculated using the lowest energy conduction band of a ZGR with $W=4b$ and $\lambda_R=0.4t$. Different  curves
correspond to different values of $k_x$ of Fig.~(\ref{band-4-.4})
}
 \label{prob-4-.4}
\end{figure}

With the expressions obtained for the wave-functions, it is possible to calculate various quantities. In particular, Fig.~(\ref{prob-4-.4}) shows the spatial probability distribution for $S^z$, the $z$-component of the spin operator defined as $<S^z>=|u_{A\up}|^2+|u_{B\up}|^2-|u_{A\down}|^2-|u_{B\down}|^2$ for the lowest energy conduction band of a ribbon with $W=4b$ and $\lambda_R=0.4t$.

The figure highlights the fact that the RSO interaction produces a clear spin polarization on the edge states of the ZGR. The non-homogeneous spin distribution across the ribbon is, however, highly dependent on the state considered. This is in contrast to the effect produced by the I-SO interaction where each state becomes spin-polarized with the same spatial spin distribution\cite{Zarea2}.

\section{Conclusions}

Graphene ribbons show unique and interesting transmission properties due to its band-structure and the pseudo-spin nature of its wave-functions \cite{Beenakker,Beenakker2}.
The relativistic nature of the description normally used makes it necessary to understand further other relativistic effects that could alter their transport properties. In this work we have investigated the consequences of one of such interactions: the Rashba spin-orbit interaction that is expected to be relevant under applied external bias voltages.
We have shown that in graphene sheets, the RSO removes the SU(2) spin degeneracy as expected while it does not open a gap in the spectrum. It does, however, introduce additional Dirac points in the Fermi surface at low energies due to crossings between valence and conduction bands. 

Because of its peculiar edge band, zigzag graphene ribbons are potential candidates for spintronic applications. The edge bands are expected to be magnetically unstable and as such to be strongly affected by SO interactions. We have shown that the RSO in particular produces states that have spin polarization and are strongly localized along the edges. These states present opposite polarization at opposite edges and the spatial spin distribution is strongly dependent on the state under consideration. Without external fields the net spin polarization of the ribbon remains null as a natural consequence of the conservation of  time reversal symmetry under the RSO interaction. However, these results suggest the possibility to obtain spin polarized currents if the states selected by an applied external voltage sustain an {\it average} non-zero spin polarization. 

\section{Acknowledgements}
We acknowledge S. E. Ulloa and G. Diniz for useful discussions. This work was partially supported by NSF under grants $N^0$ DMR 0710581 and PHY05-51164 and Ohio University BNNT funds.

\end{document}